\documentclass[12pt]{iopart}

\usepackage{epsfig}
\usepackage{graphicx}
\usepackage{amsfonts}
\usepackage{color}
\usepackage{multirow}
\usepackage{amssymb}

\begin{document}

\title[Resonance states]{Resonance states in a cylindrical quantum dot with an
external magnetic field}
\author{Alba Y Ramos and Omar Osenda}

\address{Facultad de Matem\'atica, Astronom\'{\i}a y F\'{\i}sica, Universidad
Nacional de C\'ordoba, C\'ordoba,
Argentina and IFEG-CONICET, Ciudad Universitaria, X$5016$LAE
C\'ordoba, Argentina}

\begin{abstract}
Bound and resonance states of quantum dots play a significant role in
photo-absorption processes. In this work, we analyze a cylindrical quantum dot,
its spectrum and, 
in particular, the behaviour of the lowest resonance state when a magnetic field
is applied along the symmetry axis of the cylinder. To obtain the energy and
width of the resonance we use the complex rotation method. As it is expected
the structure of the spectrum is strongly influenced by the Landau levels
associated to the magnetic field. We show how this structure affects the
behaviour of the resonance state and that the binding of the resonance has a
clear interpretation in terms of the Landau levels and the probability of
localization of the resonance state. The localization probability and the
fidelity of the lowest energy state allow to identify two different
physical regimes, a large field-small quantum dot radius regime and a small
field-large quantum dot radius, where the binding of the resonance is
dominated by the field strength or the potential well, respectively.

\end{abstract}

\pacs{73.22.-f,31.15.-p}
\maketitle

\section{Introduction}
\label{sint}

The availability of quantum devices with characteristic length of only
a few nanometer allows the implementation of a number of experimental setups
that put under test the very foundations of Quantum Mechanics. The fact that
the radius of a Landau level is of only a few nanometers for magnetic field
strengths of around 10 Tesla allows the verification of the  
Aharonov-Bohm effect in quantum rings  \cite{Bayer2003}, which
leads to
the
presence of persistent currents \cite{Mailly1993}. This persistent currents
have been measured even for one electron states \cite{Kleemans2007}. The
quantum rings are formed by only one
semiconductor material, and the fabrication of multiple concentric 
quantum rings
(up to five) can be achieved with high quality and reliability
\cite{Somaschini2009}.  

The early realization that the quantum dots presented an ideal scenario to study
transitions between electrically confined states and Landau-type magnetic
levels \cite{Sikorski1980}, originated numerous theoretical and experimental
works attempting to explain the effect of the confinement potential over the
observed spectrum. For example, Peeters and collaborators studied the spectral
properties of two-dimensional parabolic confinement potential
\cite{Peeters1990}, the effect of the confinement in the direction of the field
(or $z$ direction) \cite{Geerinckx1990,Peeters1996}, and the spectral
properties of a electron confined in an artificial molecule
\cite{Partoens1999}. Anyway, even in the case of a finite confinement potential
in the $z$ direction \cite{Peeters1996}, these studies were restricted to the
discrete spectrum of each problem.

On the other hand, the  quantum dot
has been appointed as one of the most promising implementations of a single
qubit \cite{Loss1998,Takahashi2011}. There has been a huge amount of work to
circumvent the numerous associated problems: decoherence \cite{Petta2005}, the
coupling between two qubits (to implement a two-qubit quantum gate)
in double quantum dots has been studied extensively, in
particular how it can be tuned using electric fields \cite{Kwa2009}, magnetic
fields \cite{Szafran2004}, or the effect of the confinement of the double
quantum dot in a quantum wire \cite{Zhang2008}, etc.

When dealing with applied magnetic fields, most theoretical studies on
quantum dots focus on strongly localized states, many times achieved using
always-bounding potentials, as the three dimensional harmonic potential or
impenetrable walls. The presence of a constant  magnetic field, anyway, is
equivalent to a two dimensional harmonic potential in the plane orthogonal to
the field direction. Conversely, there are far less examples of studies
considering finite potentials, in particular those whose shapes or features
allow the presence of resonance states. There is a number of reason to study
finite potentials, from charge transport situations to the effect of resonance
states in luminescent quantum dots. Bylicki and Jask\'olski
\cite{Bylicki1999}  analyzed the binding of shape resonances  through the
application of an external magnetic field. They considered an one-electron
spherical quantum dot-quantum well structure (QDQW). In this work we consider a
closely related problem with the purpose of a better understanding of the
transition from a resonance state to a bounded one. 

As has been said above, a constant magnetic field induces a two-dimensional
harmonic potential that precludes the appearance of resonances in the plane
orthogonal to the field, {\em i.e.} the loss of particles can take place
{\bf only} in the direction of the field, closing a number of decaying channels
that would be available in absence of the field. In this sense, it renders
almost irrelevant what particular shape has the binding potential, as long as
it allows the appearance of resonances in the direction of the field. So, to
avoid unnecessary complications, we consider a quantum dot model with
cylindrical symmetry whose axis has the same direction that the magnetic field,
besides we use the Effective Mass Approximation (EMA). As it is well known in
this approximation the many-body interactions of the electron trapped in the
quantum dot are reduced to a (simple) bounding potential, and all the parameters
of the Hamiltonian, mass, dielectric constants, and so on are taken as equal to
the the bulk parameters.

Despite its apparent simplicity, the calculation of the resonance states of  
one electron trapped in a bounding potential with an external magnetic field is
far from trivial, in particular for small enough strengths of the field since
the Landau levels are bunched when the strength of the field goes to zero. To
obtain the energy and width of the resonance states we employ the complex
scaling method  \cite{Moiseyev1998} that, together with a square-integrable
variational  approximation for the wave function, has been used
to analyze the bound and resonance states of two electron quantum dots
\cite{Pont2010}. Moreover, as has been shown in \cite{Pont2010}, the fidelity
of the variational eigenstates is a good tool to detect the resonance states.
In this work we use the fidelity to study the binding of a resonance and show
that it is signaled by a sharp change in the behavior of the fidelity. This
feature is consistent with the behavior observed in the fidelity when the
system experiments a quantum phase transition (in many-body models), is near
the ionization threshold or to a resonance \cite{Pont2010}.  Recently, the
application of concepts from  Quantum Information Theory, as the fidelity or
entanglement, has been very fruitful to analyze bound and resonance states in
few body systems as two-electron quantum dots
\cite{Pont2010,Ferron2009,Plastino2012,Abdullah2009}, or two-electron He-like
systems \cite{Osenda2007,Majtey2012}.

The paper is organized as follows, the quantum dot model
and the variational approximation that provides approximate
eigenvalues and eigenfunctions are presented in Section~\ref{smodel}. The
analysis of the resonance using complex exterior scaling and the influence of
the Landau levels are presented in Section~\ref{detecting} , while the 
binding process is studied, using different methods,  in
Sections~\ref{detecting},\ref{sec:localization} and \ref{sec:fidelity}. Finally,
we
summarize and discuss our results in Section~\ref{discusion}. Some rather
lengthly and technical results, mostly matrix elements are deferred to the
Appendices


\section{Model}
\label{smodel}

The bounding potential of the quantum dot is given by a piecewise function
\begin{equation}\label{eq:potential}
V(\rho,z) = \left\lbrace 
\begin{array}{lcl}
~V_1, &\quad &  \rho<a_{\rho},~\frac{a_z}{2}<|z|<\frac{a_z+b_z}{2}\\
-V_2, &\quad & \rho<a_{\rho},~|z|<\frac{a_z}{2}\\
~0, & & \rho\geq a_{\rho},~ |z|\geq\frac{a_z+b_z}{2}
\end{array}\right.
\end{equation}
{\em i.e.} the potential is a cylindrical well aligned with the z-axis, with
two potential steps at the top and the bottom of the cylinder, where the radius
of the cylinder is $a_{\rho}$, its height $a_z$, the depth of the potential
well, $-V_2$, and the height of the potential steps, $V_1$. This ensures that,
for properly chosen constants $a_z,b_z,a_{\rho}, V_1$ and $V_2$ the one electron
problem has resonance states
without external field.

Even with the introduction of the external field the angular momentum $L_z$ is
a conserved quantity, and its eigenvalues good quantum numbers. So, we focus on
states with zero angular momentum $L_z$. In this case, the Hamiltonian reads as
\begin{equation}\label{eq:hamiltoniano}
H_0= -\frac{1}{2\mu}\left(\frac{1}{\rho}\frac{\partial}{\partial\rho}\left(\rho
\frac{\partial}{\partial\rho}\right)+\frac{\partial^2}{\partial z^2}\right)
+ V(\rho,z)
\end{equation}
where $V(\rho, z)$ is given by Equation~\ref{eq:potential}, $\mu$ is the
effective mass of the electron in the semiconductor material. The Equation is
written in atomic units. If a magnetic
field $\mathbf{B}= B \hat{\mathbf{z}}$ is applied, there is a new term that must
be incorporated to the RHS of Equation~\ref{eq:hamiltoniano}, which results in
the following Hamiltonian \cite{Bylicki1999}
\begin{equation}\label{eq:hamil}
 H = H_0 + \frac{B^2}{8\mu}\rho^2.
\end{equation}

The discrete spectrum  and the resonance states of the model given by 
Eq.  (\ref{eq:hamil})  can be obtained approximately 
using square-integrable
variational functions  $\Psi_j^v$
\cite{Pont2010,Ferron2009,Bylicki2005,Kruppa1999}. So, if
$\Psi_j$ are the exact eigenfunctions of the
Hamiltonian, we look for variational approximations 

\begin{equation}\label{variational-functions}
\Psi_j \,  \simeq\, 
\Psi_j^v  \, =\, 
\sum_{i=1}^N c^{(j)}_{i}  \Phi_i 
 \, ,\;\; c^{(j)}_{i} = (\mathbf{c}^{(j)})_i 
\;\;\;\;\; ,j=1,\cdots,N \, ,
\end{equation}

\noindent where the $ \Phi_i $ must be chosen adequately,
$N$ is the
 basis set size, and the $(\mathbf{c}^{(j)})_i$ are the linear variational 
parameters of the Rayleigh-Ritz method. 

Since we are interested in null angular momentum eigenfunctions, and taking
into account the symmetries of the Hamiltonian~\ref{eq:hamil}, we choose as
basis functions
\begin{equation}
 \Phi_i(\rho,z) = \psi_n(\eta\rho) \phi_t(\nu z),
\end{equation}
where

\begin{equation}
\psi_n(\eta\rho)=\frac{1}{\sqrt{n+1}}\eta e^{-\eta\rho/2}{
L}_n^{(1)}\left(\eta\rho\right),
\end{equation}
and
\begin{equation}
 \phi_t(\nu z)=\sqrt{\frac{\nu}{2}}e^{-\nu |z|/2}{
L}^{(0)}_t\left(\nu|z|)\right),
\end{equation}
$\eta$ and $\nu$ are the non-linear variational parameters,  $L_n^{(1)}$  and 
$L_t^{(0)}$
are associated Laguerre polynomials. As has been analyzed in previous works
\cite{Pont2010,Ferron2009}, when dealing with resonance states it is convenient
to choose small values for the non-linear variational parameters. In 
particular, along
this work we use $\eta=\nu=0.01$. If $\psi_n(\eta \rho)$, where
$n=1,\ldots,N_{\rho}$, and $\phi_t(\nu z)$, where $t=1,\ldots,N_z$, then $N=N_z
N_{\rho}$.

The matrix elements of the
kinetic energy, the bounding potential and the magnetic field term are given in
the Appendix. With all these matrix elements we get a variational eigenvalue
problem
\begin{equation}
 \tilde{\mathbf{H}} \mathbf{c}^{(j)} = E^v_j  \mathbf{c}^{(j)},
\end{equation}
where the entries of the matrix $\tilde{\mathbf{H}}$ are given by
\begin{equation}\label{elemareal}
 \tilde{\mathbf{H}}_{n,t,s,m} = \left\langle \psi_n \phi_t |H| \psi_s \phi_m
\right\rangle
\end{equation}

The ample range of materials and  structures available to design self-assembled
quantum dots precludes the possibility of a very general analysis, however, in
part for comparison reasons, we use similar parameters than those used by
Bylicki in~\cite{Bylicki2008} and Bylicki and Jask\'olski in \cite{Bylicki1999}
to model a quantum dot quantum well structure made of gallium arsenide (GaAs)
composites, {\em i.e.} $a_\rho=a_z=7nm$,
$b_z=2.5nm$,
$V_1=0.37eV$, $V_2= 0.108844eV$, and $\mu=0.041m_0$. The potential well depth
is slightly different from the value used by Bylicki, but should be chosen so
 owed to the cylindrical symmetry employed in this work, conversely to the
spherical one used by Bylicki. As we will show, this set of parameters is
consistent with a resonance energy around $20meV$ with $B=0$. 

\begin{figure}
\begin{center}
\includegraphics[scale=0.5]{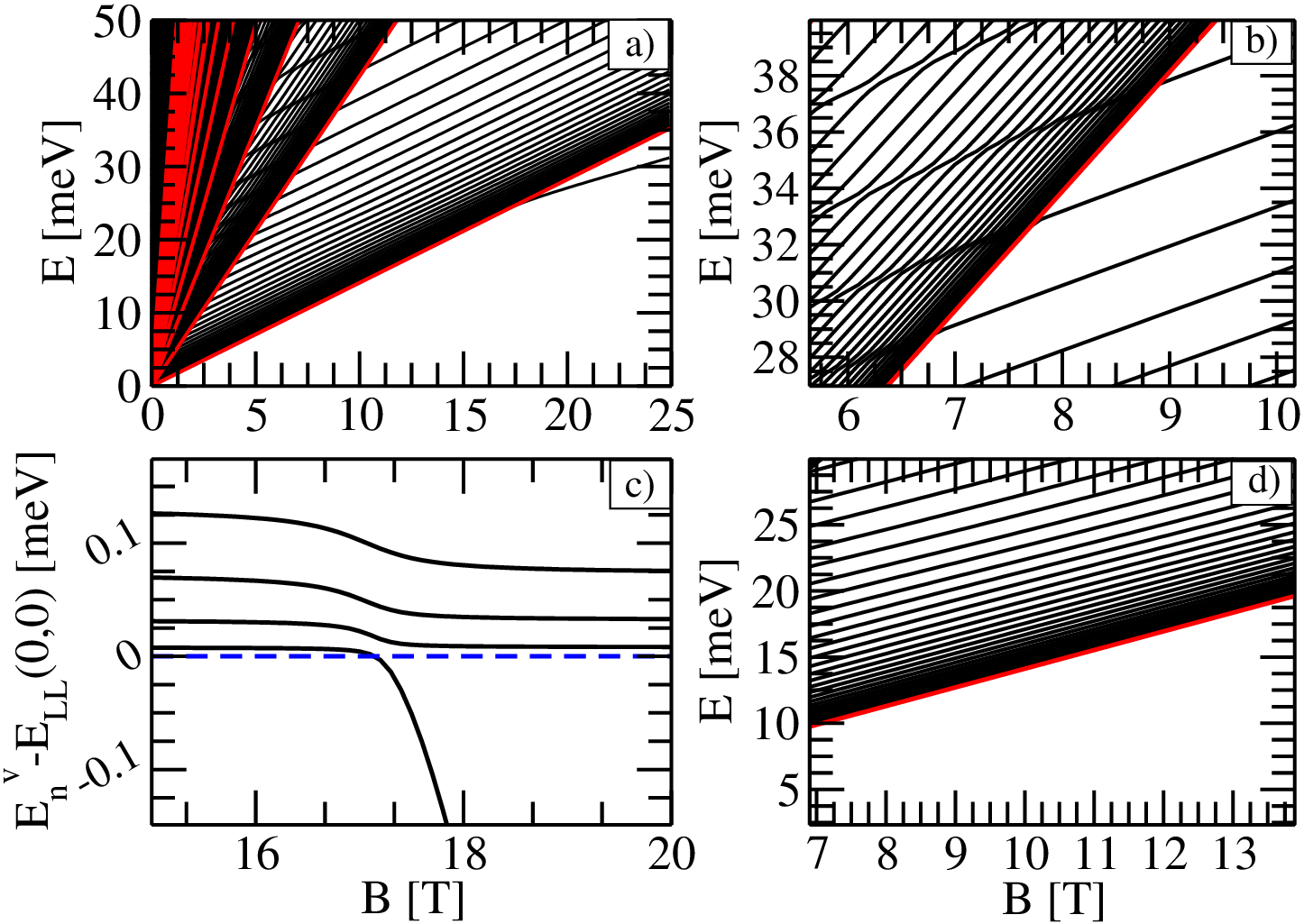}
\end{center}
\caption{\label{fig:one}(Color on-line) The four panels show details
of the variational spectrum and  the Landau levels.
a) The variational
spectrum obtained with a basis set size $N=2500$. Each black curve corresponds
to a single eigenvalue and the red curves correspond to the exact Landau
levels. b) This panel shows the zone around the second LL (red solid line). It
can be seen how the levels that cross above the second LL have several
avoided crossings before they reach the third one.
c) This panel shows the 
difference between the lowest variational eigenvalues,$E_{n}^v$, and the
first LL , $E_{\mathrm{LL}}(0,0)$, {\em vs.} the magnetic field, for
$n=1,\ldots,4$. When the
difference is smaller than zero, $E_{1}^v$ crosses the first LL as can also be
appreciated  in panel a) besides, when $E_1^v < E_{\mathrm{LL}}(0,0)$ the
corresponding eigenstate becomes localized.
d) The eigenvalues above the first LL. The figure shows clearly how the
eigenvalues are, basically parallel to the LL and accumulate above it.
}
\end{figure}

Figure~\ref{fig:one} shows the variational eigenvalues and the exact LL as
functions of the magnetic field strength $B$. The spectrum
in Figure~\ref{fig:one} was obtained using a basis set size 
$N_z=N_{\rho}=50$. As  can
be
appreciated, the eigenvalues are grouped  in sets that are bounded
between Landau levels given by $E_{\mathrm{LL}}(M,n) = \frac{B}{\mu}(|M|+n
+1/2)$,
where $M$ is the azimuthal angular momentum (in our case $M=0$), and
$n=0,1,2,\ldots$.
A pair of Landau levels delimits
a
region of the $(B,E)$ plane, in this zone the eigenvalues are parallel to the
lower Landau level, except for the appearance of avoided crossings. This
particular feature is shown in Figure~\ref{fig:one} b). 

As can be seen very clearly from Figure~\ref{fig:one}d), the variational
eigenvalues have the tendency to accumulate above the Landau levels. This
feature is similar to the eigenvalue accumulation  observed above the continuum
threshold of a two-electron quantum dot \cite{Geerinckx1990,Ferron2009}. When a
magnetic field
is applied, each Landau level
works as the bottom of a continuum. Another salient feature of the spectrum
appears for large enough magnetic fields: one isolated eigenvalue with
lower energy than the lowest Landau level. From a physical point of view, the
origin of this state can be understood as follows: for intermediate values of
the magnetic fields the wave function of the electron looks like an harmonic
oscillator wave function in the $(\rho,\phi)$ plane and as a free particle in
the $z$ direction. The spatial extent of the wave function on the plane is
roughly equivalent to the radius of a Landau level, that is larger than the
radius (in the plane) of the quantum dot. When the strength of the magnetic
field is increased the radius of the lowest Landau level becomes smaller and
smaller reaching, at some point, a size similar to the radius of the quantum
dot, at this point the state becomes localized. As we will show, the mechanism
of localization can be quantified and strongly influences the behaviour of the
resonance states whose energy lies near the localization point, in particular
this mechanism is responsible of the binding of the resonance for large enough
magnetic field \cite{Bylicki1999}, we will be back to this point in
Section~\ref{sec:localization}. 

The similarities between the behaviour of the
spectrum analyzed in Figure~\ref{fig:one} near the point where $E_1^v \sim
E_{\mathrm{LL}}(0,0)$,  and the spectrum of a two-electron quantum dot near the
ionization threshold are striking. So, it comes as no surprise that the
resonance states of both models also show some similarities, as we will show
later on.

\section{Detecting the resonance states using complex exterior scaling}
\label{detecting}

The calculation of the energy and width of resonance states offers a number of
challenges that
repeatedly leads to the formulation of new methods. Among the most widely
used methods  can be mentioned the complex scaling (or complex dilation) method
\cite{Moiseyev1998}, the complex absorbing potential method \cite{Sajeev2009}
and  the density of states method \cite{Mandelshtam1993}. Each one of these
has its advantages and drawbacks. Because we are dealing with a
piecewise potential we resort to the exterior complex  scaling (see, for
example, Reference~\cite{Moiseyev1998}). This method is particularly appealing
in our case
since for $B\neq 0$, the complex scaling should be applied only to the $z$
variable because this is the only direction available to the electron to get
away from the bounding potential. The {\em exterior} of the method refers to the
exterior of the region where the potential is not zero so, for $B\neq 0$ the
exterior complex scaling asks that
\begin{equation}\label{eq:complex-scaling}
 z \longmapsto \left\lbrace
\begin{array}{lcl}
 z^{\prime} & \mbox{if} & |z| \leq \frac{a_z+b_z}{2} , \\
e^{i\theta} z^{\prime}  & \mbox{if} & |z| \geq \frac{a_z+b_z}{2} 
\end{array}
\right. ,
\end{equation}
where $z$ is the coordinate to be complex scaled, and $\theta$ is
the rotation angle.
  The 
complex scaling turns the 
 resonance states into normalizable states (with a different
norm) which can be analyzed with the variational approach usually employed
in Hermitian problems, {\em i.e.} the energy and width of the resonance state
can be obtained approximately as an isolated complex eigenvalue of a finite
(complex) Hamiltonian, if $H(\theta)$ is the complex scaled Hamiltonian, then 
we get a complex variational eigenvalue
problem
\begin{equation}
 \tilde{\mathbf{H}}(\theta) \mathbf{d}^{(j)} = E^v_j(\theta)  \mathbf{d}^{(j)},
\end{equation}
where the entries of the matrix $\tilde{\mathbf{H}}(\theta)$ are given by
\begin{equation}\label{elemacomplex}
 \tilde{\mathbf{H}}_{k,n,l,m}(\theta) = \left\langle \psi_k \phi_n |H(\theta)|
\psi_l \phi_m
\right\rangle .
\end{equation} 
Another advantage of the complex scaling method is given by the similarities
between the  matrix
elements in Equation~\ref{elemacomplex}
with the ones calculated in Equation~\ref{elemareal}.

\begin{figure}
\begin{center}
\includegraphics[scale=0.3]{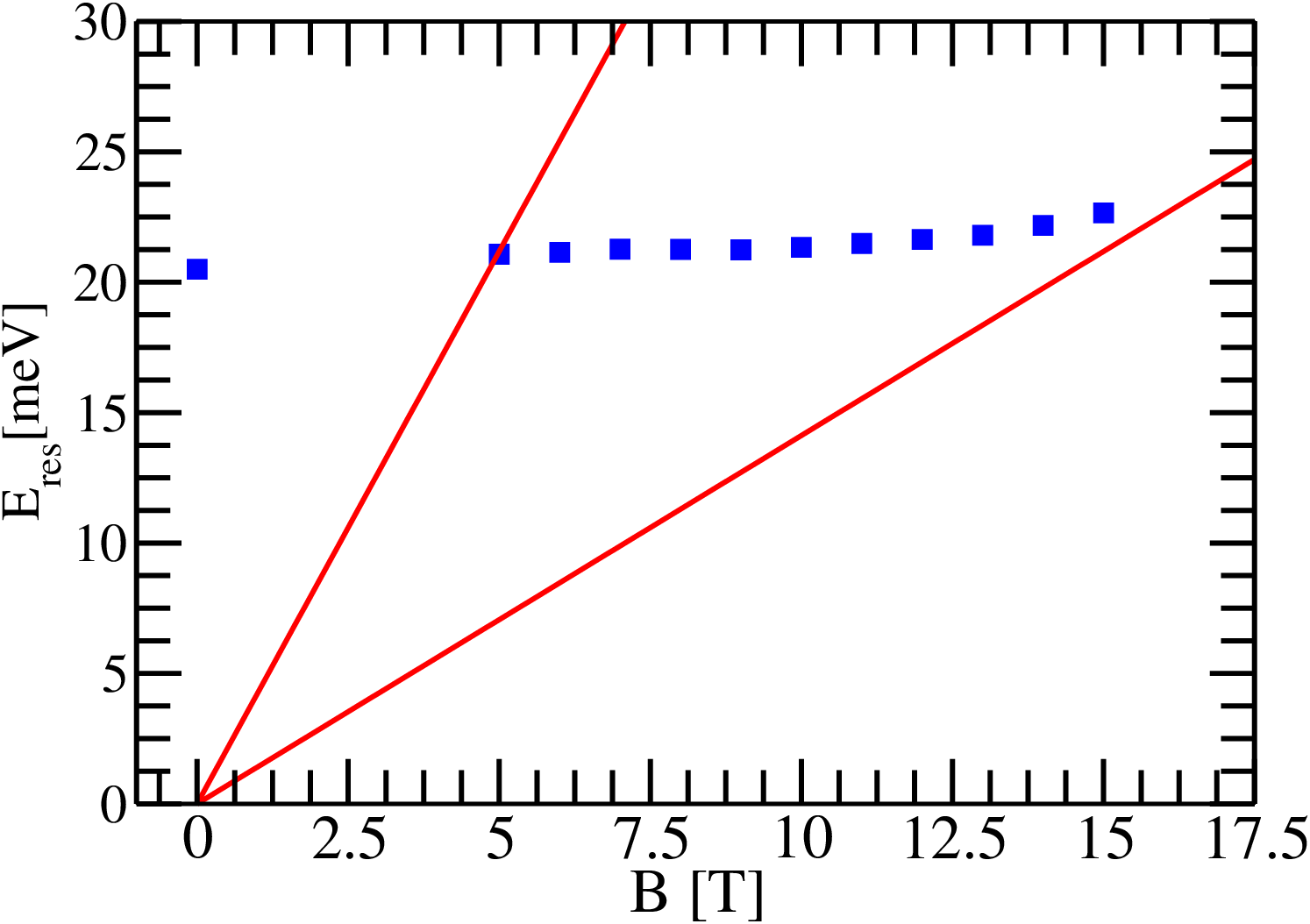}
\includegraphics[scale=0.3]{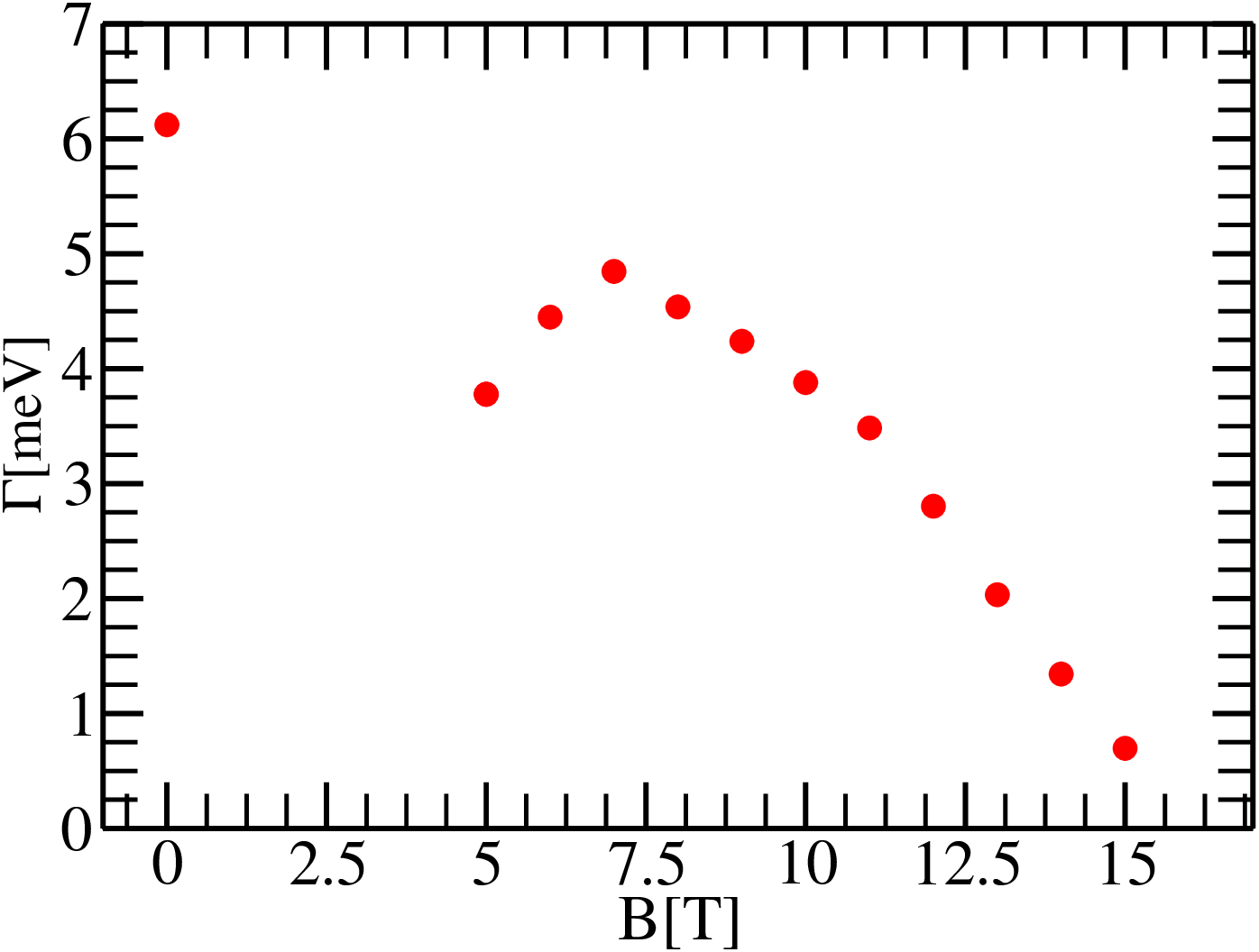}
\end{center}
\caption{\label{fig:two} The real and  imaginary parts of the
resonance eigenvalue $E(\theta)$ {\em vs} the magnetic field strength. a) The
solid square (blue) dots correspond to the energy of the resonance and the 
solid lines correspond to the two lowest Landau level energies. The value for
$B=0$ was obtained performing the complex rotation in the coordinates $\rho$
and $z$. b) The width of the resonance {\em vs} the magnetic field. As the
magnetic field increases its value the width of the resonance goes to zero,
signaling its binding. 
}
\end{figure}

Figure~\ref{fig:two} shows the behaviour of the energy and width of the lowest
resonance as a function of the magnetic field strength. The binding of the
resonance can be clearly observed, {\em i.e.} the width of the resonance drops
to zero when the magnetic field strength increases its value.
To obtain the data shown in
Figure~\ref{fig:two}a) and b)
we looked for the best value obtainable from the method,
the complex rotation was performed for different values of $\theta$ and the
best approximation corresponds to the stationary points of the
$\theta$-trajectory \cite{Moiseyev1998}.

\begin{figure}
\begin{center}
\includegraphics[scale=0.3]{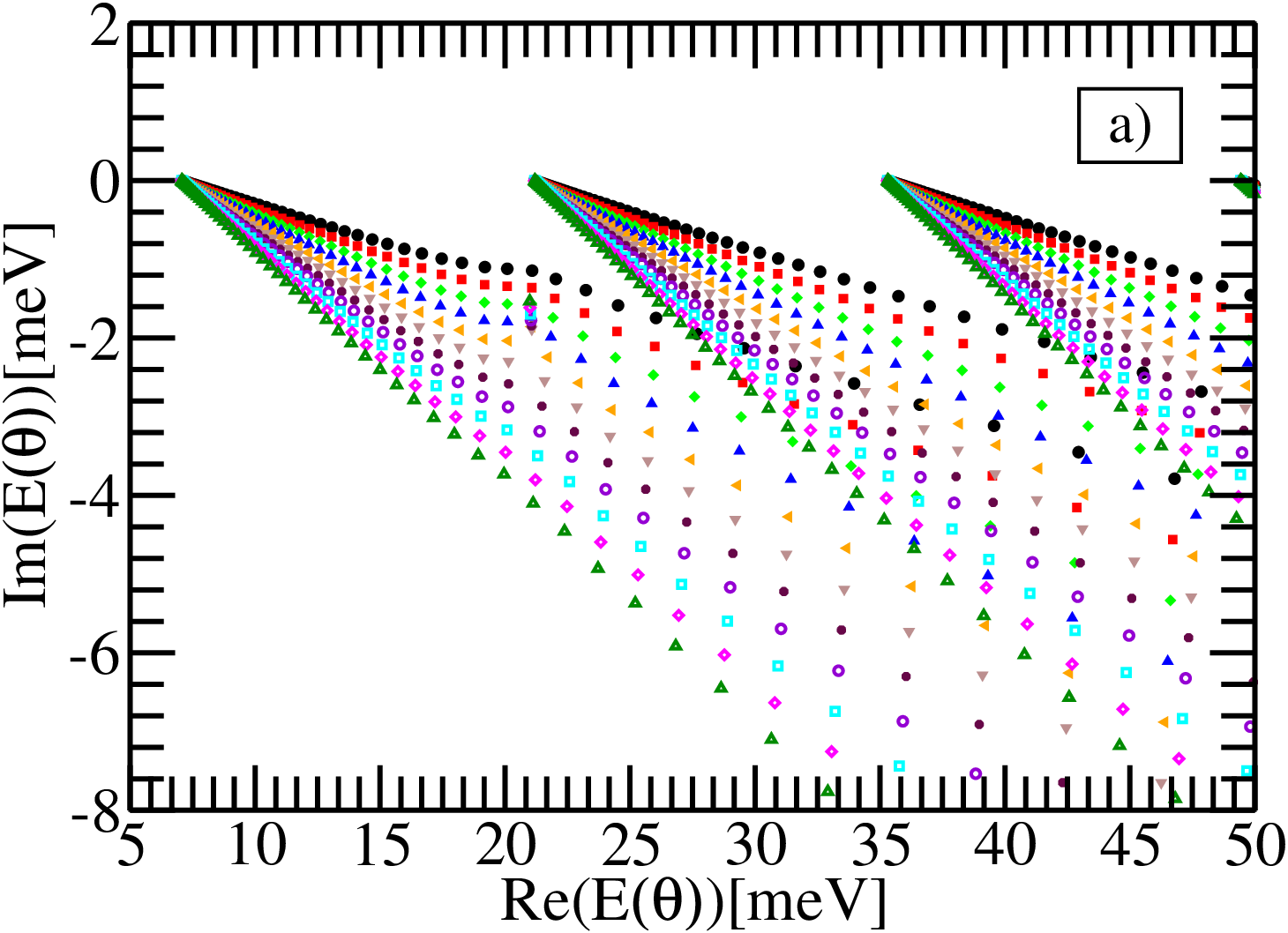}
\includegraphics[scale=0.3]{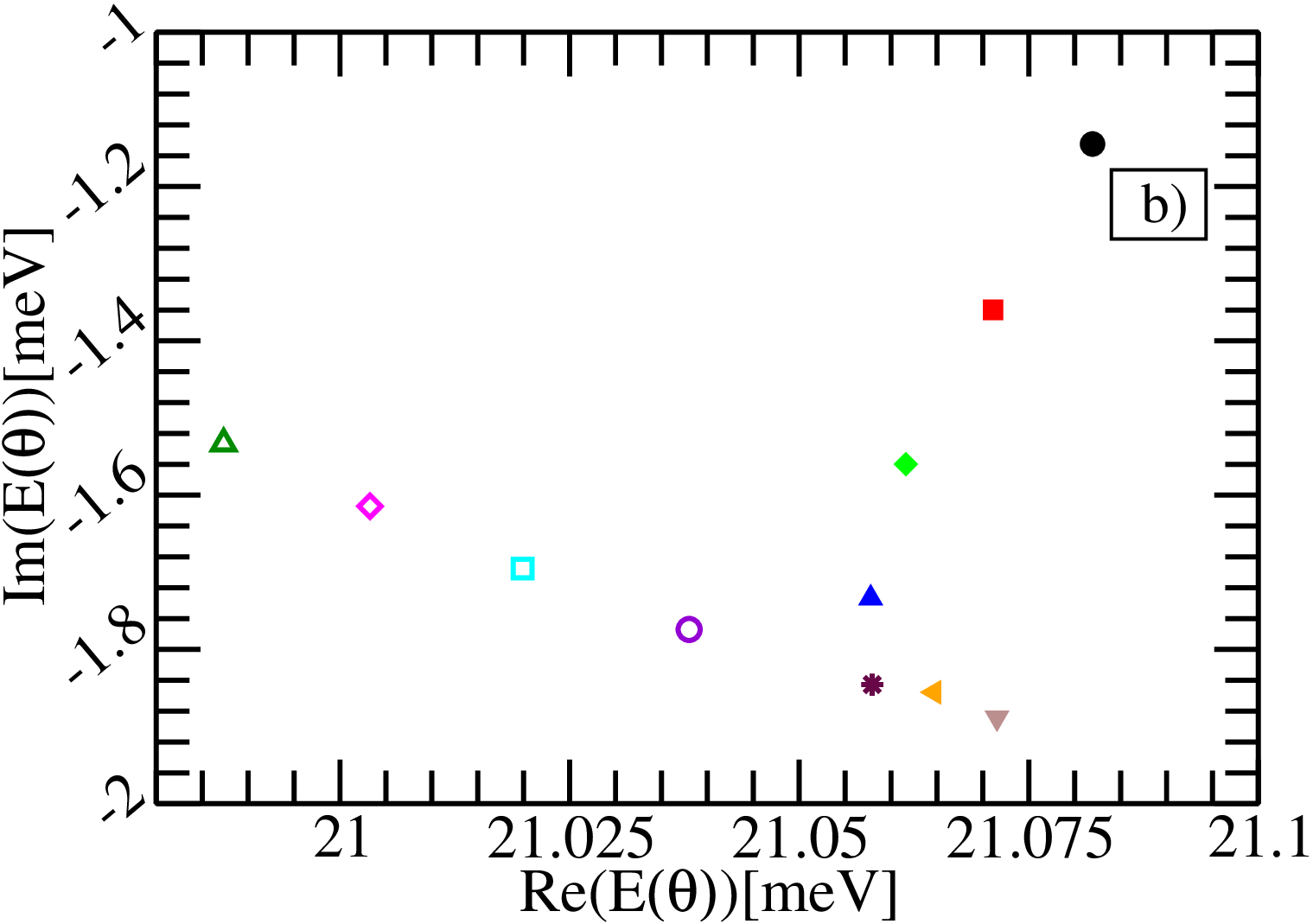}
\caption{\label{fig:three}(Color on-line) a) The imaginary part 
{\em vs} the real part of
the complex rotation eigenvalues . All the eigenvalues
were calculated for $B=5T$ and different rotation angles.
From top to bottom,
the eigenvalues correspond to $\theta=0.1$ (black
\fullcircle), $\theta=0.11$ (red {\color{red}\fullsquare}), $\theta=0.12$ (green
{\color{green} $\blacklozenge$}), up to $\theta=0.2$. The three fan-like sets
of data are related to the first three LL's. Is is clear that the the
leftmost fan and the central one  start to overlap around $27 \, \mbox{meV}$.
For $B<5T$ the overlap grows larger making more difficult to look for the
resonance data. b) The $\theta$ trajectory of the resonance eigenvalue. The
symbols correspond to the angles shown in panel a). The eigenvalues that form 
the $\theta$ trajectory can also be
observed in panel a), they lie near the bump that presents the leftmost fan
around $2 \, \mbox{meV}$
}
\end{center}
\end{figure}

The exterior
complex scaling methods works nicely for intermediate values of the magnetic
field. Anyway, if $B \leq 5\mathrm{T}$, or $B\geq 16\mathrm{T}$, the convergence
of the method is,
at least, questionable, see Figure~\ref{fig:three}. For small fields the 
method struggles to provide a
reliable value for the resonance eigenvalue because, as can be appreciated from
Figure~\ref{fig:two}, for $B\approx 5\mathrm{T}$ the resonance ``enters'' in the
continuum above the second Landau level (LL). In this region the approximate
eigenvalues that enter from the region above the first Landau level ``collide''
with the eigenvalues that lie between the first and second LL. The successive
collisions, and the corresponding avoided crossings, can be well appreciated in
Figure~\ref{fig:one}b).

The binding of the resonance, {\em i.e.} that the width of the resonance
becomes zero for large enough field strengths can not be understood only
studying the spectrum. In the next Section, we will introduce a quantity that
will allows us to study the associated eigenstates. As we will show, the
analysis of the  eigenstate corresponding to the isolated eigenvalue that
appears below the first LL, see Figure~\ref{fig:one}a), gives a physical picture
of the binding process.

In what follows, and up to the end of this Section, we want to focus in the
results of the complex rotation method between Landau levels.
The scenario between LL's can be better appreciated in Figure~\ref{fig:three}.
The Figure shows the complex spectrum obtained when the complex rotation is
performed accordingly  with Equation~\ref{eq:complex-scaling} for different
values of $\theta$. In this Figure it is clear why the method is termed complex
rotation, the continuum part of the spectrum now lies in the complex plane over
straight lines, the angle between the lines and the real axis equal to
$2\theta$. When $B\neq 0$, the data for different $\theta$'s form 
``hand-fans'' of data, {\em i.e.} sets of straight lines with a common origin
in the real axis. Each one of these sets can be associated to a single LL, the
leftmost hand-fan corresponds to first LL, the middle one to the second LL, and
the rightmost to the third LL. Again, this structure can be attributed to role
that each LL plays as the bottom of a continuum of levels. Effectively, the
distance between the origins of each set of lines, over the real axis, is
exactly equal to the distance between the LL at the magnetic field strength
considered. As the magnetic field strength decreases, the data of the different
fans overlaps making  very difficult to distinguish the eigenvalues 
corresponding to the resonance.

Figure~\ref{fig:three}b) shows the trajectory in the complex plane of the
eigenvalue associated to the resonance state. As can be appreciated, the
eigenvalues are much less scattered in energy than in width. This feature is
typical of the complex rotation method. 

Above the second LL the complex rotation method should be complemented with
some {\em ad hoc} assumptions to identify the points that belong to
the $\theta$-trajectory. The main assumption is rather physical: the complex
rotated eigenvalues can be approximated by $E_{LL}+ \tan{(2\theta)}\, E_n$, {\em
i.e.} they form a straight line in the complex plane that form an angle of
$2\theta$ with the $x$-axis, $E_n$ is a variational eigenvalue. This
approximation works nicely {\bf except} near a resonance. The second assumption
is similar: a given complex eigenvalue $E_{\nu}(\theta)$ should not change too
much if $\theta$ is changed, say
\begin{equation}
 |E_{\nu}(\theta+\delta \theta) - E_{\nu}(\theta)| \leq \delta\theta
|E_{\nu}(\theta)|,
\end{equation}
where $\delta \theta$ is small enough. Applying these assumptions and
discarding complex eigenvalues that jump from a hand-fan to another (see
Figure~\ref{fig:three}a)) it is feasible to obtain stabilized values for
$E_{res}$ above the second LL.

\section{Detecting the resonance states using Localization Probability}
\label{sec:localization}

As we will show in this Section, 
the binding of the resonance for large enough magnetic field is a consequence
of the strong localization experienced by the wave function corresponding to 
the lowest eigenvalue. As a matter of
fact, the localization  allows to follow the
resonance through the spectrum. To quantify the localization of an approximate
eigenfunction $\Psi_j^v $, we calculate the probability that the electron is
localized in the potential well,
\begin{equation}
 P_j = \int_{-\frac{a_z}{2}}^{\frac{a_z}{2}} \int_0^{a_{\rho}} \left|
\Psi_j^v(\rho,z)\right|^2 \, d\rho \, dz \; ,
\end{equation}
where $P_j$ is the probability attributable to the localization of $\Psi_j^v$.

\begin{figure}
\begin{center}
\includegraphics[scale=0.28]{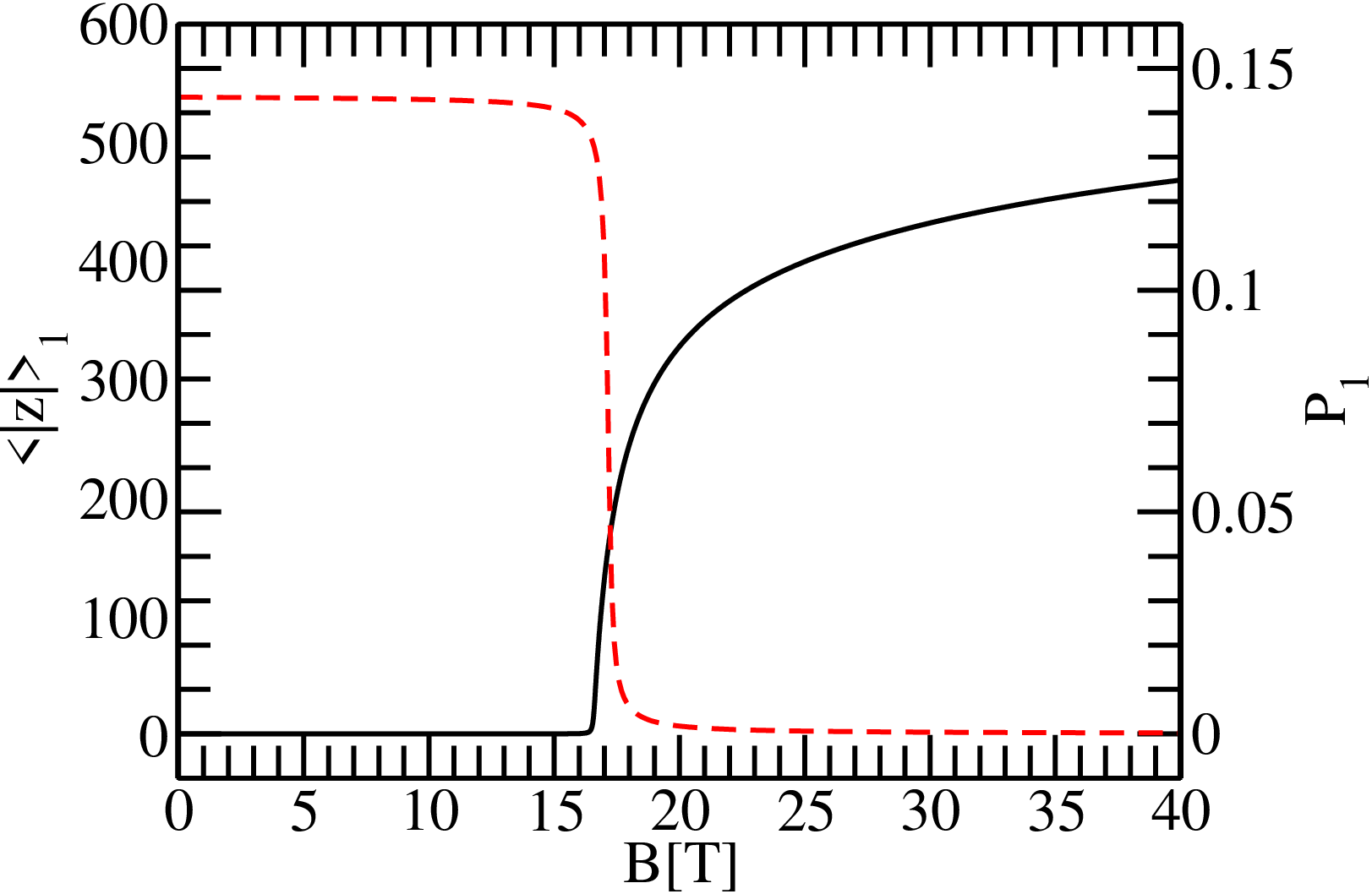}
\includegraphics[scale=0.28]{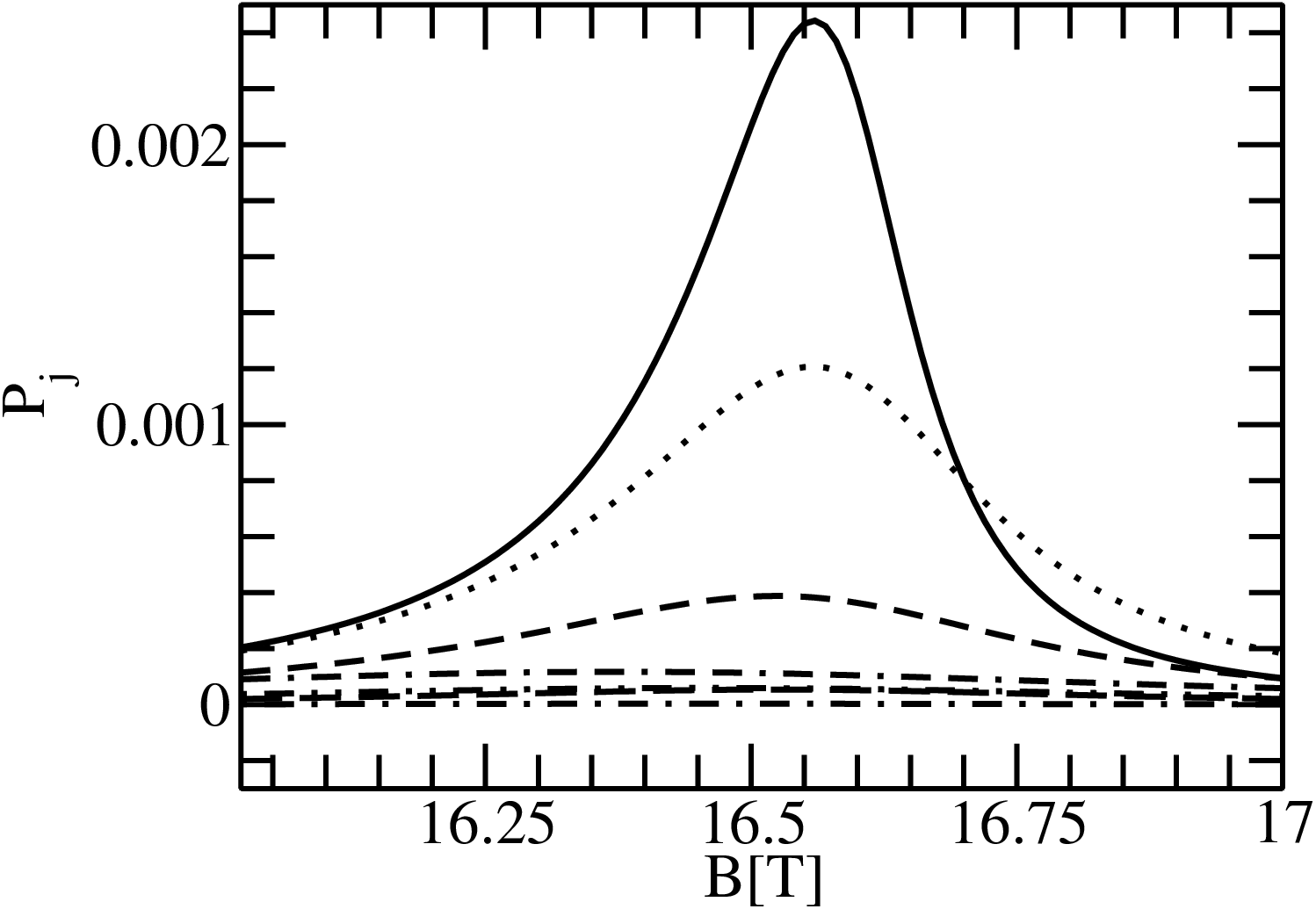}
\end{center}
\caption{\label{fig:four} The localization probability {\em vs}
the magnetic field strength. The left panel shows $P_1$ (black solid
line), {\em i.e.}  the
probability corresponding to the lowest variational eigenvalue, and
$\left\langle |z|\right\rangle_1$ (red dashed line). The scale of the right
ordinate axis corresponds to $P_1$, while the scale of the left one corresponds
to $\left\langle
|z|\right\rangle_1$. The right panels shows, from top to bottom, $P_2$ (solid
line), $P_3$
(dotted-line), $\ldots$, $P_8$. The localization probabilities show a peak
where the corresponding eigenvalue crosses near the resonance, after and before
the peak the eigenstate is extended.
}
\end{figure}

Figure~\ref{fig:four}a) shows the probability $P_1$ as a function of the
magnetic field strength for the lowest variational eigenvalue, while
Figure~\ref{fig:four}b) shows the probability $P_j$ for $j=2,3,\ldots,8$. The
curves
are shown in different panels because of their respective scale.

It is worth to mention that despite the variational eigenvalues  do not show any
sudden changes near $B\sim 17$T, except for the lowest one when it
crosses the first LL, the probability of localization into the well shows a
well defined maximum for a large number of eigenvalues. 

For small values of the magnetic field, the probability of localization
$P_1 \lesssim 10^{-3}$ and is a very smooth function of $B$. This behaviour
changes abruptly near $B\sim 17\, \mathrm{T}$, and for larger values of
$B$ the localization probability grows up to values larger than $0.1$, see
Figure~\ref{fig:four}a). It is
clear that this behaviour is not compatible with an extended wave function.
Since we are approximating the exact solutions of the eigenvalue problem with
square-integrable functions, the value of the localization probability can not
drop to zero, in the continuum region and far away from a resonance state, the
variational eigenfunctions resemble a plane wave that is extended in a
cylindrical spatial region with radius $R\sim 1/\eta$ and length $L\sim 1/\nu$,
where $\eta$ and $\nu$ are the non-linear variational parameters. For this 
reason, in the
continuum region and far away from a resonance, we estimate  the localization
probability as $P_1\sim a_z a_{\rho}\eta \nu\sim 10^{-3}$. 

The localization of  $\Psi_1^v$ and   can be further analyzed
using the expectation value 
\begin{equation}
\left\langle |z|\right\rangle_1 = \int_{-\frac{a_z}{2}}^{\frac{a_z}{2}}
\int_0^{a_{\rho}} \left|
\Psi_1^v(\rho,z)\right|^2 \, |z| d\rho \, dz \; , 
\end{equation}
where $|z|$ is the absolute value of $z$.
Consistently
with the behaviour observed for the localization probability, the expectation
value $\left\langle |z|\right\rangle_1$ is very large for $B<17 \, \mathrm{T}$
and drops its value around two orders of magnitude for $B>17 \, \mathrm{T}$,
see Figure~\ref{fig:four}a). We conclude that the wave function of the lowest
variational eigenvalue becomes bounded. For the example that we are
analyzing, $\left\langle |z|\right\rangle_1\sim 540$ nanometers, for $B=0$, and
drops to $\left\langle |z|\right\rangle_1\sim 3$ nanometers for $B\sim 18\,
\mathrm{T}$. This last value is consistent with the longitudinal dimension of
the quantum dot. Besides, as shown in
Figure~\ref{fig:one}a), the lowest eigenvalue is isolated from the continuum for
$B>17 \, \mathrm{T}$. 

The scenario depicted above leads us to the following conclusion, the resonance
above the first LL is triggered by the ``collision'' of the bounded state,
corresponding to the lowest eigenvalue, with the continuum above the first LL.
The avoided crossings originated by this collision can be clearly appreciated
in Figure~\ref{fig:one}c), and reinforces the interpretation that the binding
of the resonance is produced when a bounded state appears for large enough
magnetic field.
In this sense, the resonance state behaves as a shape resonance and the radial
potential owed to the magnetic field is the one changing its shape. The
crossing of the lowest eigenvalue with the first LL corresponds, obviously, to
the continuum threshold. Finally, as is the case in shape resonances, when the
lowest eigenvalue is bounded and isolated the resonance width is zero, which
explains the binding of the resonance for large enough fields.

The behaviour of the localization probabilities $P_j$, with $j>1$ is quite
different. These $P_j$ are not monotonically increasing functions of $B$ in
contradistinction with $P_1$, instead they shown a more or less pronounced
maximum
for some value of $B$, see Figure~\ref{fig:four}b). The height of the maximum
of a given $P_j$ depends on the sharpness of the avoided crossing experienced
by the corresponding eigenvalue. Sharper avoided crossings lead to higher
maxima, smoother ones lead to lower maxima. This is consistent with the
numerical spectrum obtained, since the avoided crossing of a low lying
eigenvalue is always sharper than the avoided crossing of a higher eigenvalue.

We interpret that the peak
in a given localization probability appears when the corresponding eigenvalue is
a reasonable approximation for the resonance energy at that particular value of
$B$. In other words, $ E_{res}(B_j^{peak})\approx E_j^v(B_j^{peak})$, where
$B_j^{peak}$ is the value of the magnetic field where $P_j$ attains its
maximum. The $E_j^v(B_j^{peak})$ allow to track down the resonance states from
the localization point up to the second LL, above the second LL the spectrum
shows a multiple-continua region, generating multiple avoided crossings due to
two-continua interaction that this method can not distinguish from the avoided
crossings due to resonances. The $E_j^v(B_j^{peak})$ are shown in
Figure~\ref{fig:seven}, where they are compared to the resonance energies
obtained using the other methods discussed in this work.

\section{Detecting the resonance states using the Fidelity}
\label{sec:fidelity}

For a given quantum state $\psi$, that depends on a parameter $\lambda$, a
measure of how much it changes when the parameter is varied is given by the
fidelity, $\mathcal{F}$, which is defined as
\begin{equation}\label{eq:fidelity}
\mathcal{F}_{\Delta \lambda}(\lambda)= \left| \langle \psi(\lambda-\Delta
\lambda),
\psi(\lambda+\Delta \lambda)
\rangle\right|^2 ,
\end{equation}
where $\Delta \lambda$ is a small variation of the parameter.

\begin{figure}
\begin{center}
\includegraphics[scale=0.3]{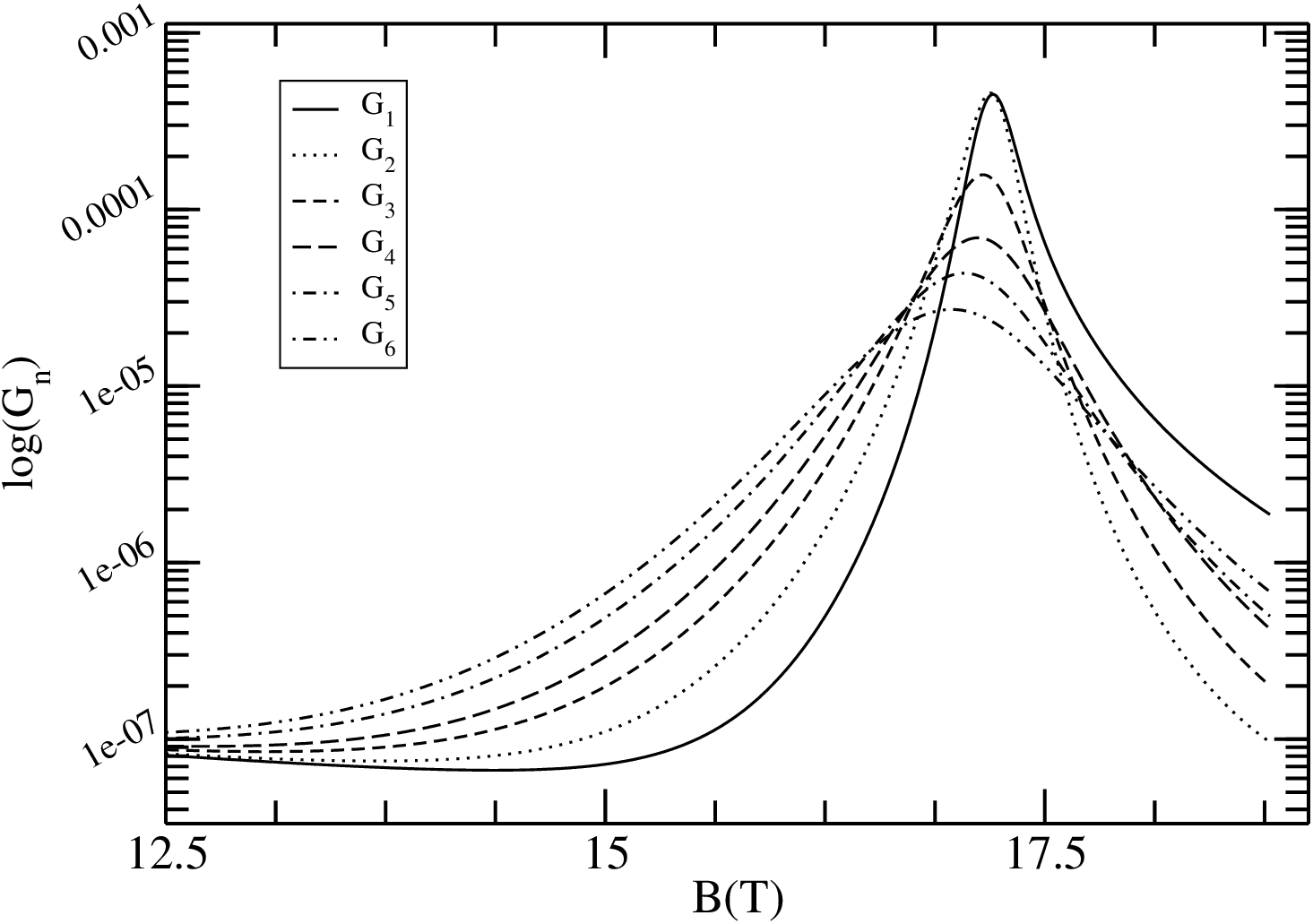}
\end{center}
\caption{\label{fig:five} The function $\mathcal{G}_n$ {\em vs} the magnetic
field. The figure shows the function for the first six variational eigenvalues,
$\mathcal{G}_1$ , $\mathcal{G}_2$, $\ldots$, $\mathcal{G}_6$, solid (\full), 
dotted (\dotted), short-dashed (\dashed), long-dashed (\longbroken), dot-dashed
(\chain) and  double dot-dashed (\dashddot) lines respectively. The peak where
each curve attains its maximum value is clearly appreciable. $B_n$ is given by
the abscissa of the peak (see the text).
}
\end{figure}

The fidelity has been extensively used to characterize the analytical
properties of quantum states near a quantum phase transition in quantum spin
chains models \cite{Zanardi2006},  quantum phases of matter
\cite{Zanardi2009},or bound and resonance states in atomic or quantum dot models
\cite{Pont2010}. In two-electron quantum dots, it has been shown that the
fidelity of the  approximate variational eigenstates detect the resonance
states and allow to calculate approximately its energy \cite{Pont2010}.
To achieve this, the fidelity of many eigenstates should be calculated as a
function of the external parameter that drives the system  from bound to
unbounded states. In this work, this parameter is the magnetic field strength.

From its definition, Equation~\ref{eq:fidelity}, it is clear that most of the
time $\mathcal{F}_{\Delta \lambda} \approx 1$, except for very special cases. On
the other hand the fidelity should drop to zero if the system experiments a
rather sudden change \cite{Zanardi2006}. For this reason, often it is convenient
to
study the function
\begin{equation}
 \mathcal{G}_n(\lambda) = 1 - \left| \langle \Psi_n^v(\lambda-\Delta
\lambda),
\Psi_n^v(\lambda+\Delta \lambda)
\rangle\right|^2 .
\end{equation}

Figure~\ref{fig:five} shows the behaviour of the $\mathcal{G}_n$ function for
$n=1,2,\ldots, 6$, as function of the magnetic field strength. Each curve has a
more or less well defined peak for a given value of the magnetic field, $B_n$.
The values $E_n(B_n)$, {\em i.e.} the value of the variational eigenvalues at
their respective peaks of the function $\mathcal{G}_n$, give a very good
approximation for the resonance energy at the points $B_n$.  This way to obtain 
a estimation for the resonance energy
does not allow to obtain it for
arbitrary magnetic field strength values since the values $B_n$ are not chosen
at will, they  are obtained from the fidelity data and depend on the
basis set size, the non-linear variational parameters, the basis functions 
used, and so on,
this has been pointed out previously in Reference~\cite{Pont2010}. Nevertheless
the method provides another
tool to analyze resonance states. The fidelity method works best when the width
of the resonance is not too wide, so it is to be expected that it will be more
precise near the localization point. 

\begin{figure}
\begin{center}
\includegraphics[scale=0.3]{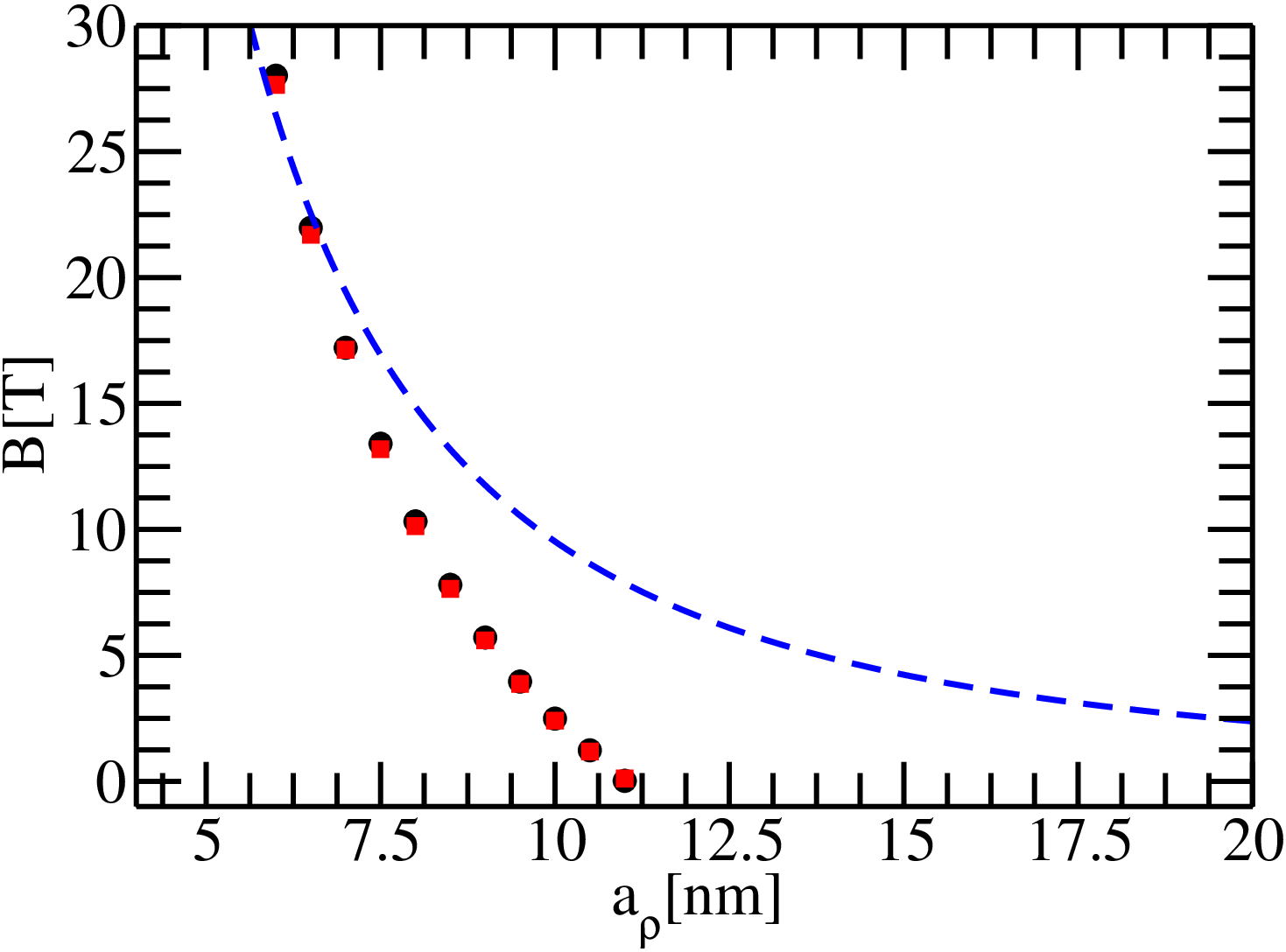}
\end{center}
\caption{\label{fig:six}(Color on-line) The critical fields $B_p(a_{\rho})$ and
$B_{\mathcal{F}}(a_{\rho})$ {\em vs} the quantum dot radius. The localization
probability critical field (red squared dots, {\color{red}\fullsquare}) and
the fidelity critical field (solid black dots, \fullcircle) data is shown for
several quantum dot radius. The (blue) dashed curve correspond to the radius of
the lowest LL as a function of the magnetic field. It is clear that for small
quantum dot radius
the localization takes place when the cylindrical wave function enters into the
cylindrical QD. For larger QD's radius the localization is dominated by the
quantum well potential and not by the magnetic field.
}
\end{figure}

Before presenting the results of the localization and fidelity methods with
respect to the resonant states, we want to stress the relationship between both
quantities.  Let us call $B_p(a_{\rho})$ the critical value of the magnetic
field such that the localization probability of the first variational
eigenvalue becomes noticeable for a given quantum dot radius $a_{\rho}$, and
$B_{\mathcal{F}}(a_{\rho})$ the magnetic field value such that the function
$\mathcal{G}_1$ attains its maximum value as a function of $B$.
Figure~\ref{fig:six} shows the
values of both quantities, $B_p(a_{\rho})$ and $B_{\mathcal{F}}(a_{\rho})$, for
several values of the quantum dot radius $a_{\rho}$. The agreement between both
critical quantities is striking.

\begin{figure}
\begin{center}
\includegraphics[scale=0.3]{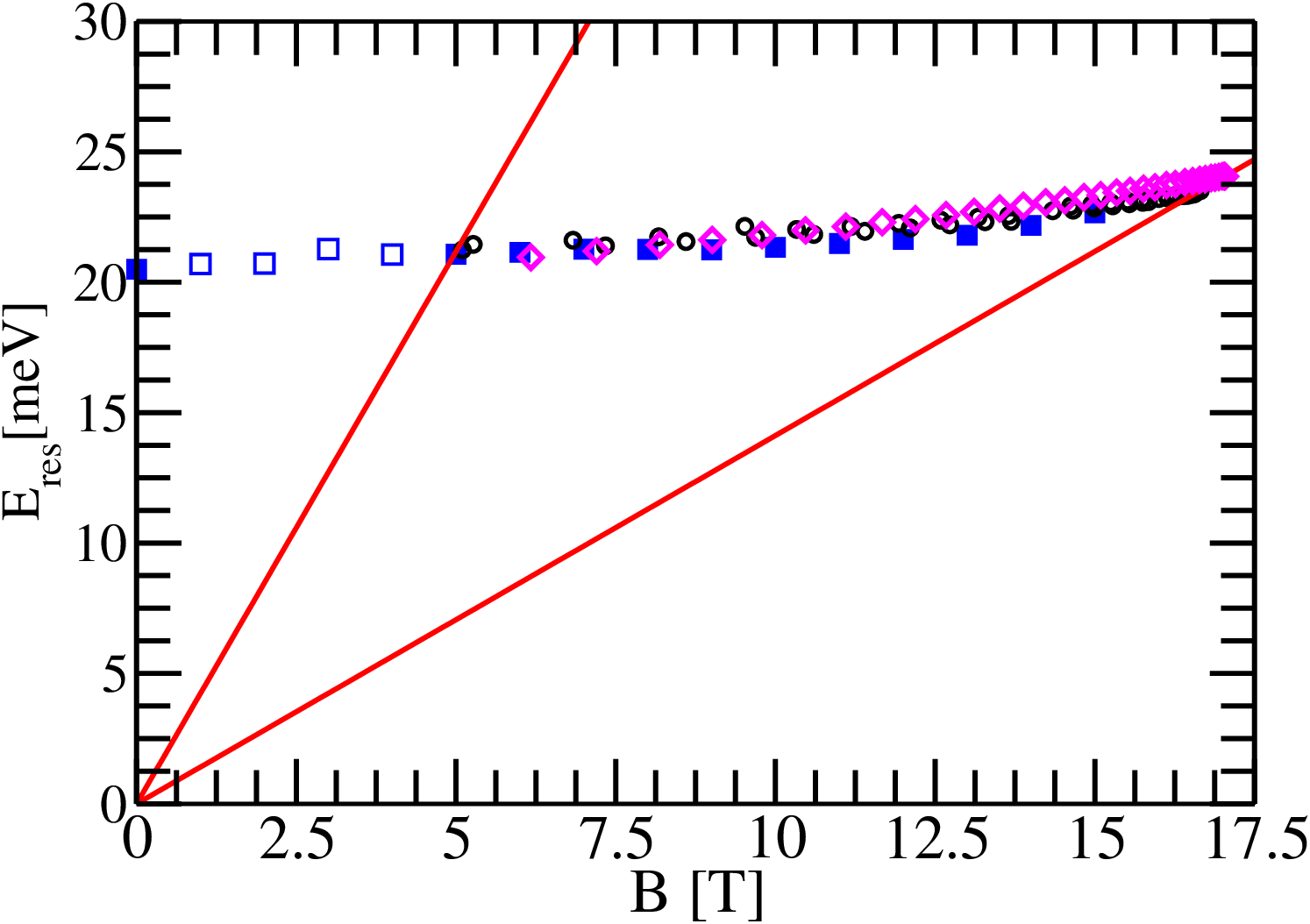}
\end{center}
\caption{\label{fig:seven} The resonance energy $E_{res}$ calculated using the
three methods depicted in the text, complex exterior scaling, localization
probability, fidelity, the corresponding values are shown using blue square
dots ({\color{blue}\fullsquare}), black circles (\opencircle), and magenta
diamonds ({\color{magenta} \opendiamond}). The energy of the resonance in the
region above the second LL can be obtained using a modified version of the
complex rotation method and is shown using blue open squares ({\color{blue}
\opensquare}) .
}
\end{figure}

At this point, we can summarize our results in Figure~\ref{fig:seven}. The
Figure shows the energy of the resonant state as a function of $B$, as it is
obtained from the complex rotation, fidelity and localization methods. 

\begin{figure}
\begin{center}
\includegraphics[scale=0.3]{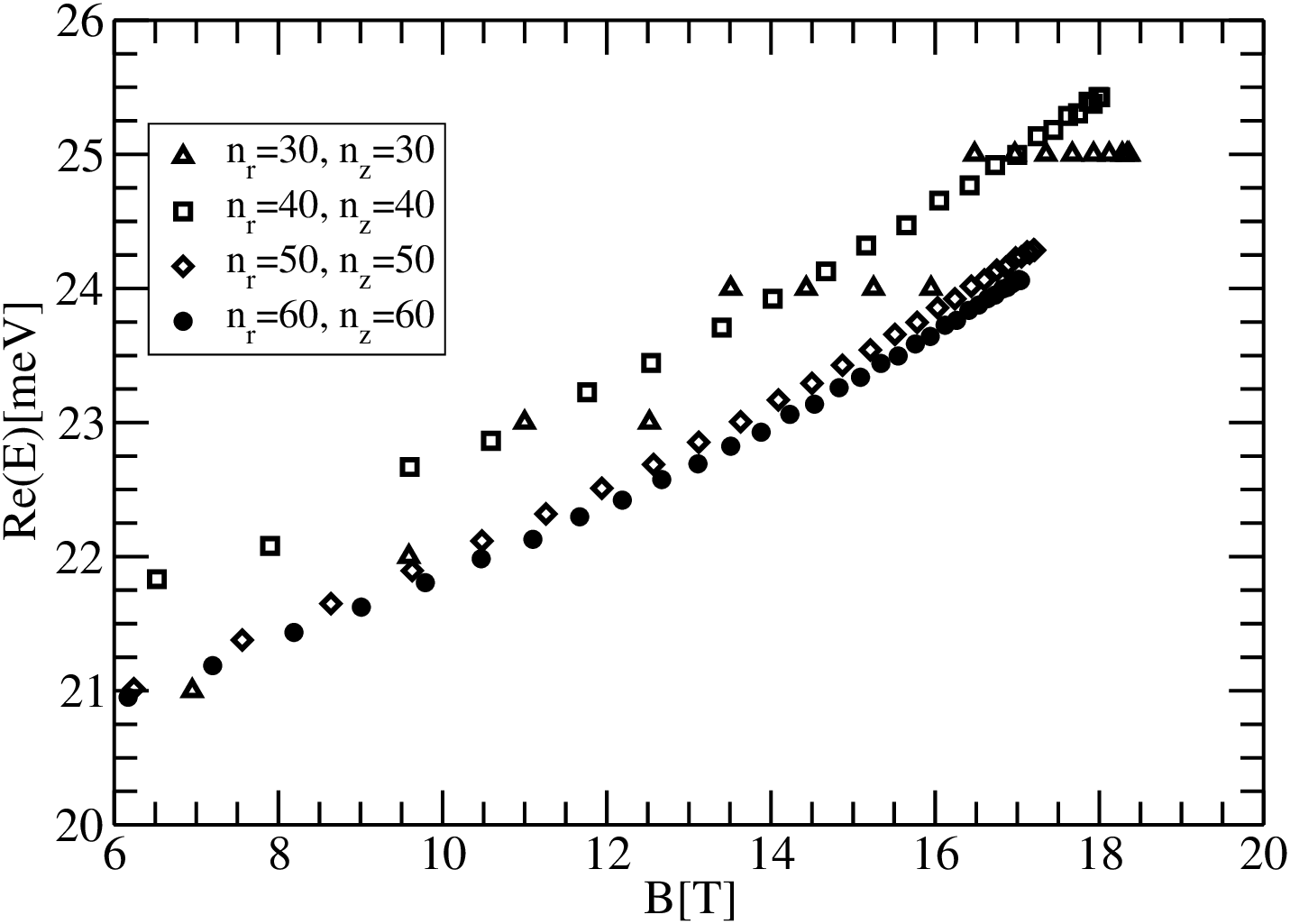}
\end{center}
\caption{\label{fig:sevenp} $E_n(B_n)$ {\em vs} the magnetic field strength.
The figure shows the data obtained using different basis set sizes,
$N=30\times30$, triangle up open dots; $N=40\times40$, square open dots;
$N=50\times50$, diamond open dots; and $N=60\times60$, solid dots.
}
\end{figure}

From Figure~\ref{fig:seven} is rather clear that both the localization method
and the fidelity 
are
able to follow the resonance from the localization point until the second
Landau level. Anyway, a word of caution is necessary here. The complex rotation
method implies a stabilization procedure that, when properly used, gives an
idea about the accuracy of the results obtained. This is not true for the
Fidelity method. The best recipe to obtain stabilized results from the Fidelity
method implies picking the largest possible basis set size, $N$, with
$N_z=N_{\rho}$. Otherwise, the convergence of the Fidelity data, {\em i.e} the
values obtained for the resonance's energy from the Fidelity, is not uniform.
Figure~\ref{fig:sevenp} shows the behaviour of the resonance energy obtained
following the Fidelity method with different basis set sizes. It is worth to
mention that the resonance energies obtained with $N=900$ and $N=3600$ 
differ in less than 5$\%$

Again, as is the case with the localization probability method to find
estimations for the resonance energy, the fidelity method works properly as
long as the resonance state is well isolated, {\em i.e.} the only avoided
crossings present in a given region of the spectrum should be one associated
to the resonance. As has been said previously, above the second LL there is a
very large number of avoided-crossings for each eigenvalue. owed to a
multiple-continua scenario. Each $\mathcal{G}_n$ has a peak associated to
every single avoided-crossing, making extremely hard to decide which peak
corresponds to the resonance state.

\section{Discussion and Conclusions}
\label{discusion}

The external magnetic field precludes  the escape of the electron on the
$(\rho,\phi)$ plane, for this reason the quantum dot bounding potential
consists of a potential well and a potential step, to ensure the presence of
resonance states. In this sense, the problem has several characteristic
lengths, the quantum dot radius, the Landau levels radii, the length of the
quantum dot along the $z$ axis, and so on. The two lengths that come into play
varying the magnetic field strength, for the set of parameters considered in
this work, are the quantum dot radius and the lowest LL radius. The potential
well depth and the other parameters of the QD were chosen to ensure that for
$B=0$ there were no bound states. 

The localization probability is able to track down the resonance states from
the localization point up to the second LL. Above the second LL the eigenvalues
have too many avoided crossings, because of the multiple-continua interacting in
this region of the spectrum,
rendering the method useless. The many avoided
crossings also prevent the use of the fidelity to detect the resonance above the
second LL. Moreover, despite that the width of the resonance state analyzed is
not small when compared with the resonance energy, the fidelity method provides
a good estimation of the last. Anyway, some precaution must be exercised to
obtain accurate and stable results.

Finally, both critical fields, $B_p(a_{\rho})$ and
$B_{\mathcal{F}}(a_{\rho})$, show two clearly distinguishable regimes, for
small quantum dot radius their values are given by the lowest Landau level
radius, so both are proportional to $1/a_{\rho}^2$. For large enough
quantum dot radius both critical fields show a different behaviour and,
apparently, both are proportional to $(\alpha -\beta a_{\rho})^2$, where
$\alpha$ y $\beta$ are constants. The extent of the small field regime can be
tuned changing the parameters of the bounding potential,
Equation~\ref{eq:potential},
extending or reducing it. The critical behaviour of the eigenvalues in the
transition region between the large field-small quantum dot radius and the
small field-large quantum dot radius will be analyzed elsewhere.


\section*{Acknowledgments}

We would like to acknowledge SECYT-UNC, CONICET,
and MinCyT C\'ordoba for partial financial support of this
project. We would like to thank Dr. Pablo Serra for helpful comments and the
critical reading of this manuscript.


\appendix

\section{Some matrix elements}

It is convenient to separate the kinetic energy matrix elements in two
contributions, one corresponding to the radial coordinate,

\begin{eqnarray}
\left\langle \psi_n\left|T_{\rho}\right|\psi_s\right\rangle &=& \left\langle
\psi_n\left|-\frac{1}{2\mu}\nabla_r^2\right|\psi_s\right\rangle \nonumber \\
&=&\frac{\eta}
{\mu\sqrt{(n+1)(s+1)}}\left(\frac14T_1+\frac12\left(T_2+T_3\right)+T_4\right)
\end{eqnarray}
and the other corresponding to the $z$ coordinate,
\begin{eqnarray}
\left\langle \phi_t\left|T_{z}\right|\phi_r\right\rangle &=& \left\langle
\phi_t\left|-\frac{1}{2\mu}\nabla_z^2\right|\phi_r\right\rangle
\nonumber \\
&=&\frac{\eta}
{2\mu\sqrt{(n+1)(s+1)}}\left(\frac14T_{1z}+\frac12\left(T_{2z}+T_{3z}\right)+T_{
4z}\right)
\end{eqnarray}
where
\begin{eqnarray}
 T_1&=&(n+1)\delta_{n,s},\nonumber\\
 T_2&=&\sum_{p=0}^n\sum_{q=0}^{s-1}\frac{(-1)^{p+q}(n+1)!(s+1)!(p+q+1)!}{
(n-1-p)!
(2+p)!p!(s-q)!(1+q)!q!},\\
 T_3&=&\sum_{p=0}^{n-1}\sum_{q=0}^s\frac{(-1)^{p+q}(n+1)!(s+1)!(p+q+1)!}{(n-p)!
(1+p)!p!(s-1-q)!(2+q)!q!},\nonumber\\
 T_4&=&\sum_{p=0}^{n-1}\sum_{q=0}^{s-1}\frac{(-1)^{p+q}(n+1)!(s+1)!(p+q+1)!}{
(n-1-p)!
(2+p)!p!(s-1-q)!(2+q)!q!}\nonumber\\
\end{eqnarray}
while
\begin{eqnarray}
 T_{1z}&=&\delta_{t,r},\nonumber\\
 T_{2z}&=&\sum_{d=0}^t\sum_{f=0}^{r-1}\frac{(-1)^{t+r}t!~r!(d+f)!}{(t-d)!
(d!)^2(r-1-f)!(1+f)!f!},\\
 T_{3z}&=&\sum_{d=0}^{t-1}\sum_{f=0}^r\frac{(-1)^{t+r}t!~r!(d+f)!}{(t-1-d)!
(1+d)!(r-f)!(f!)^2},\nonumber\\
 T_{4z}&=&\sum_{d=0}^{t-1}\sum_{f=0}^{r-1}\frac{(-1)^{t+r}t!~r!(d+f)!}{(t-1-d)!
(1+d)!d!(r-1-f)!(1+f)!f!}.\\
\end{eqnarray}

The matrix element of the bounding potential can be factorized owed to its
piecewise character and using that there is a potential barrier on the $z$
direction. We get that 
\begin{equation}
\left\langle
\psi_n \phi_t\left|V(\rho,z)\right|\psi_s \phi_r\right\rangle=-V_2\,
I_{V2}+V_1\, I_{V1}\delta_{n,s},
\end{equation}
the barrier term is, obviously, proportional to $V_1$, while the term
proportional to $V_2$ corresponds to the matrix element of the potential well.
The matrix elements $I_{V2}$ and $I_{V1}$ are given by
\begin{eqnarray}
 I_{V2}&=&\frac{1}{\sqrt{(n+1)(s+1)}}\sum_{p=0}^n\sum_{q=0}^s\frac{(-1)^{p+q}
(n+1)!
(s+1)!~I_{V21}}{(n-p)!(1+p)!p!(s-q)!(1+q)!q!}\nonumber\\
&\times&\sum_{d=0}^t\sum_{f=0}^r\frac{(-1)^{d+f}t!~r!~I_{V22}}{
(t-d)!(d!)^2(r-f)!(f!)^2}\\
 I_{V1}&=&\frac{1}{\sqrt{(n+1)(s+1)}}\sum_{d=0}^t\sum_{f=0}^r\frac{(-1)^{d+r}
t!~r!~I_{V11}}
{(t-d)!(t!)^2(r-f)!(f!)^2}
\end{eqnarray}
and 
\begin{equation}
 I_{V21} = (p+q+1)!-e^{-\eta
a}\sum_{k=0}^{p+q+1}\frac{(p+q+1)!(a\eta)^{p+q+1-k}}
{(p+q+1-k)} ,
\end{equation}
\begin{equation}
 I_{V22} = (d+f)!-e^{-\nu
a/2}\sum_{g=0}^{d+f}\frac{(d+f)!}{(d+f-g)}\left(\frac{\nu a}{2}\right)^{d+f-g} ,
\end{equation}
\begin{eqnarray}
 I_{V11}&=&\sum_{k=0}^{d+f}\frac{(d+f)!}{(d+f-k)!} \nonumber \\
 && \times \left(e^{-a\nu/2}\left(\frac{
a\nu}{2}\right)^{d+f-k}
-e^{\frac{-(a+b)\nu}{2}}\left(\frac{(a+b)\nu}{2}\right)^{d+f-k}\right).
\end{eqnarray}

The matrix element of the magnetic field term reads as
\begin{eqnarray}
 \left\langle
\psi_n\left|H_c\right|\psi_s\right\rangle&=&\frac{B^2}{8\mu\eta^2\sqrt{
(n+1)(s+1) } }
\nonumber \\
&& \times
\sum_{
p=0
}^n
\sum_{q=0}^s\frac{(-1)^{p+q}(n+1)!(s+1)!(p+q+3)!}{(n-p)!(1+p)!p!(s-q)!(1+q)!q!}
\end{eqnarray}
%


\end{document}